# Twist-induced Out-of-plane Ferroelectricity in Bilayer Hafnia

*Jian Huang,*[1*] *Gwan Yeong Jung,*[2] *Pravan Omprakash,*[1] *Guodong Ren,*[1] *Xin Li,*[3] *Du Li,*[4]

*Xiaoshan Xu,*[3] *Li Yang,*[4] *and Rohan Mishra*[2,1,4*]

[1] Institute of Materials Science and Engineering, Washington University in St. Louis, St. Louis, MO 63130, USA

[2] Department of Mechanical Engineering and Material Science, Washington University in St. Louis, St. Louis, MO 63130, USA

[3] Department of Physics and Astronomy, University of Nebraska, Lincoln, NE 68588, USA

[4] Department of Physics, Washington University in St. Louis, St. Louis, MO 63130, USA

*Corresponding author: J. H. (h.jian@wustl.edu) and R. M. (rmishra@wustl.edu)

ABSTRACT: Ferroelectric $HfO_2$ is a promising candidate for next-generation memory devices due to its CMOS compatibility and ability to retain polarization at nanometer scales. However, the polar orthorhombic phase responsible for ferroelectricity is metastable and requires extrinsic stabilization, complicating integration with silicon. Here, we predict that twist-induced symmetry breaking in bilayer 1T-$HfO_2$ enables robust, switchable out-of-plane (OOP) polarization. First-principles density functional theory shows that monolayer 1T-$HfO_2$, cleaved from the (111) surface of cubic $HfO_2$, is dynamically stable. Twisting two aligned monolayers forms a moiré superlattice that breaks interlayer symmetry and yields bistable OOP polarization. At a twist angle of 7.34°, the bilayer exhibits ~16 $\mu C/cm^2$ polarization, primarily from polar displacements within AB stacking domains. The polarization can be switched via interlayer sliding with a low energy barrier (~8 meV per formula unit) and coercive field (~0.2 V/nm), enabling electric-field tunability in atomically thin ferroelectric devices.

The discovery of ferroelectricity in hafnia ($HfO_2$),[1] a material used as the dielectric layer in silicon-based transistors, has opened promising avenues for the realization of CMOS-compatible nonvolatile memory devices capable of directly integrating logic and storage functions.[2, 3] In contrast to conventional perovskite ferroelectrics such as $PbTiO_3$ and $BaTiO_3$, $HfO_2$ demonstrates scale-free ferroelectricity, retaining switchable polarization even at thicknesses of a few nanometers.[2, 4, 5] However, its orthorhombic polar phase ($Pca2_1$) is metastable as it does not appear in the bulk phase diagram,[6] and requires stabilization through extrinsic factors such as doping and strain.[7-10] Consequently, thin films of polar hafnia are often accompanied by defects and coexisting paraelectric phases that limit their potential.[11] Furthermore, the orthorhombic polar phase of hafnia exhibits hard ferroelectric behavior, with coercive fields nearly an order of magnitude higher than those of conventional perovskite ferroelectrics, thereby reducing energy efficiency.[11] Thus, despite its intrinsic compatibility with silicon and persistent ferroelectricity in ultrathin films, the utility of $HfO_2$ is limited by the need for complex phase-stabilization strategies and its characteristically large coercive field. These limitations call for the exploration of alternative approaches to achieve robust ferroelectricity with low coercivity in low-dimensional $HfO_2$ for the seamless integration of logic and memory in CMOS devices.

A new paradigm for realizing ferroelectricity in low dimensions involves manipulating the stacking order in bilayers of van der Waals (vdW) materials, giving rise to sliding and moiré ferroelectricity.[12-14] These distinct forms of ferroelectric order have been experimentally demonstrated in a variety of two-dimensional (2D) materials, including graphene,[15] hexagonal boron nitride ($h$-BN),[16] and transition metal dichalcogenides (TMDs).[17] Sliding and moiré ferroelectricity in vdW materials mitigate the issues arising from the epitaxial mismatch of the film and the substrate, such as the formation of defects and secondary phases in hafnia. vdW integration of layers having different functionalities also opens up new avenues for the development of next-generation nanoelectronics and optoelectronic devices with multifunctionalities.[18-20] However, sliding and moiré ferroelectricity in reported vdW systems typically yield modest out-of-plane (OOP) polarization, as the effect originates from subtle interlayer charge redistribution across weak vdW interfaces. Consequently, the polarization magnitude is limited to only a few tenths of $\mu C/cm^2$.[21] Furthermore, in all the reported bilayer moiré ferroelectric systems, the OOP polarization alternates between upward and downward orientations, leading to a net zero macroscopic OOP polarization.[22] In such moiré ferroelectrics, a net OOP polarization often only emerges under the application of an external electric field.[23, 24] To overcome the limited OOP polarization, recent efforts have focused on exploring stacking-dependent ferroelectricity in emerging 2D systems such as 2D oxides.[25-27] Unlike conventional 2D materials, these systems often exhibit soft-phonon-based ferroelectric instability, which, combined with the non-negligible residual stresses in membranes, makes their ferroelectric properties highly sensitive to stacking configurations and interlayer interactions.[28, 29] Recent studies have reported the emergence of topological polar textures with sizeable polarization magnitudes in twisted perovskites membranes.[30, 31] Nevertheless, robust and switchable OOP polarization in bilayer systems remains to be demonstrated.

In this Letter, we introduce bilayers of the 1T phase of $HfO_2$ as a promising platform for realizing large and switchable macroscopic OOP polarization. In contrast to sliding and moiré ferroelectricity, we find that the polarization in twisted bilayer 1T-$HfO_2$ — that has been predicted to adopt a layered structure — originates from the coupling between the local stacking order of the bilayer and the soft-phonon-induced polar atomic displacements within individual layers. In contrast to sliding ferroelectricity reported in other bilayer 2D materials, we show that interlayer sliding alone is insufficient to generate macroscopic polarization in bilayer 1T-$HfO_2$.[21, 32] Instead, relative twisting between the layers creates distinct stacking domains that modify interlayer interactions and selectively activate soft-phonon instabilities within AB domains, giving rise to robust OOP polarization. This twist-induced distortion of the vertical stacking order stabilizes a bistable polar state with a low switching barrier, resulting in a maximum predicted OOP polarization of 16 μC/cm$^2$ at a twist angle of 7.34°, approaching the theoretically reported polarization of the ferroelectric $Pca2_1$ phase of $HfO_2$ (~50 μC/cm$^2$).[33] Furthermore, the polarization can be reversibly switched between two states via interlayer sliding with a low energy barrier of ~8 meV/formula unit (f.u.), offering a novel route for electric-field control of the polarization states. These findings suggest that twisted bilayer 1T-$HfO_2$ provides a feasible and scalable platform for harnessing the intrinsic compatibility of $HfO_2$ with Si while achieving robust and switchable OOP polarization.

## RESULTS AND DISCUSSION

We first investigate the feasibility of accessing a monolayer of 1T-$HfO_2$ and assess its thermodynamic stability. Monolayer 1T-$HfO_2$ can be derived by cleaving the (111) surface of the cubic phase ($Fm\bar{3}m$), as illustrated in Figure 1a,b. The resulting structure comprises a central layer of Hf atoms sandwiched between two layers of O atoms, with each Hf atom octahedrally coordinated by six O atoms—three positioned above and three below the Hf plane.[34] Notably, the Hf–O bond length in monolayer 1T-$HfO_2$ is reduced relative to its bulk counterpart, decreasing from 2.20 Å in the $Fm\bar{3}m$ phase to 2.10 Å. The bond contraction is consistent with trends observed in other monolayer 1T phase transition metal oxides, such as 1T-$CrO_2$ and 1T-$ZrO_2$,[35, 36] suggesting enhanced bonding interactions due to reduced dimensionality.[37]

To evaluate the feasibility of exfoliating a monolayer, we calculated the cleavage energy ($E_{\mathrm{CE}}$) and surface energy ($E_{\mathrm{SE}}$) of various bulk $HfO_2$ phases in different surface orientations:[38]

$$E_{\mathrm{CE}} = \frac{E_{\mathrm{slab,as-cut}} - E_{\mathrm{bulk}}\frac{N_{\mathrm{slab}}}{N_{\mathrm{bulk}}}}{A_{\mathrm{slab}}}, \tag{1}$$

$$E_{\mathrm{SE}} = \frac{E_{\mathrm{slab,opt}} - E_{\mathrm{bulk}}\frac{N_{\mathrm{slab}}}{N_{\mathrm{bulk}}}}{2A_{\mathrm{slab}}}, \tag{2}$$

where $E_{\mathrm{slab,as-cut}}$ and $E_{\mathrm{slab,opt}}$ denote the total energies of the as-cleaved and relaxed slabs, respectively. $E_{\mathrm{bulk}}$, $A_{\mathrm{slab}}$, $N_{\mathrm{slab}}$ and $N_{\mathrm{bulk}}$ represent the energy of the bulk phase, the surface area, and the number of atoms in the slab and the bulk, respectively. Surface energy serves as the thermodynamic driving force for surface reconstructions, and minimal reconstructions are a prerequisite for the stability of a monolayer.[38] Ideally, in the absence of surface reconstructions, the surface energy is expected to be half of the cleavage energy ($E_{\mathrm{slab,as-cut}} = E_{\mathrm{slab,opt}}$). When reconstructions occur during relaxation, the surface energy further decreases, indicating such surfaces are less favorable for experimental realization compared to those exhibiting minimal reconstructions.[38] The cleavage and surface energies for various $HfO_2$ phases and surfaces are presented in Figure 1c, where the diagonal dashed line represents surfaces without reconstructions ($E_{\mathrm{SE}} = \frac{1}{2}E_{\mathrm{CE}}$). We note that most surfaces are away from the diagonal, suggesting those surfaces are less accessible. Experimentally reported surfaces are highlighted with red edges and are found to lie close to the diagonal line. Importantly, the (111) surface of the cubic phase exhibits the lowest cleavage and surface energies, and negligible surface reconstructions, suggesting high accessibility in terms of surface stability.[39, 40]

The thermodynamic stability of monolayer 1T-$HfO_2$ was further assessed by calculating the energy above the convex hull ($E_{\mathrm{hull}}$), as shown in Figure S1, where comparisons are made with various $HfO_2$ phases and experimentally reported monolayer oxides and dichalcogenides. Although the $E_{\mathrm{hull}}$ of monolayer 1T-$HfO_2$ is higher than that of the bulk phase of $HfO_2$, it remains within the amorphous limit, and is comparable to experimentally reported 2D metal oxides and chalcogenides, suggesting the feasibility of experimental realization.[41] Phonon dispersion calculations, shown in Figure 1d, reveal the absence of modes with imaginary frequencies across the Brillouin zone, confirming the dynamic stability of monolayer 1T-$HfO_2$. Furthermore, the electronic band structure, presented in Figure 1e, indicates that monolayer 1T-$HfO_2$ is a wide-gap semiconductor with an indirect bandgap of 4.9 eV at the PBE level, which is beneficial for maintaining polarization by suppressing intrinsic leakage currents. Thus, monolayer 1T-$HfO_2$ emerges as a feasible candidate for potentially achieving polarization at the nanoscale through manipulation of the stacking order.

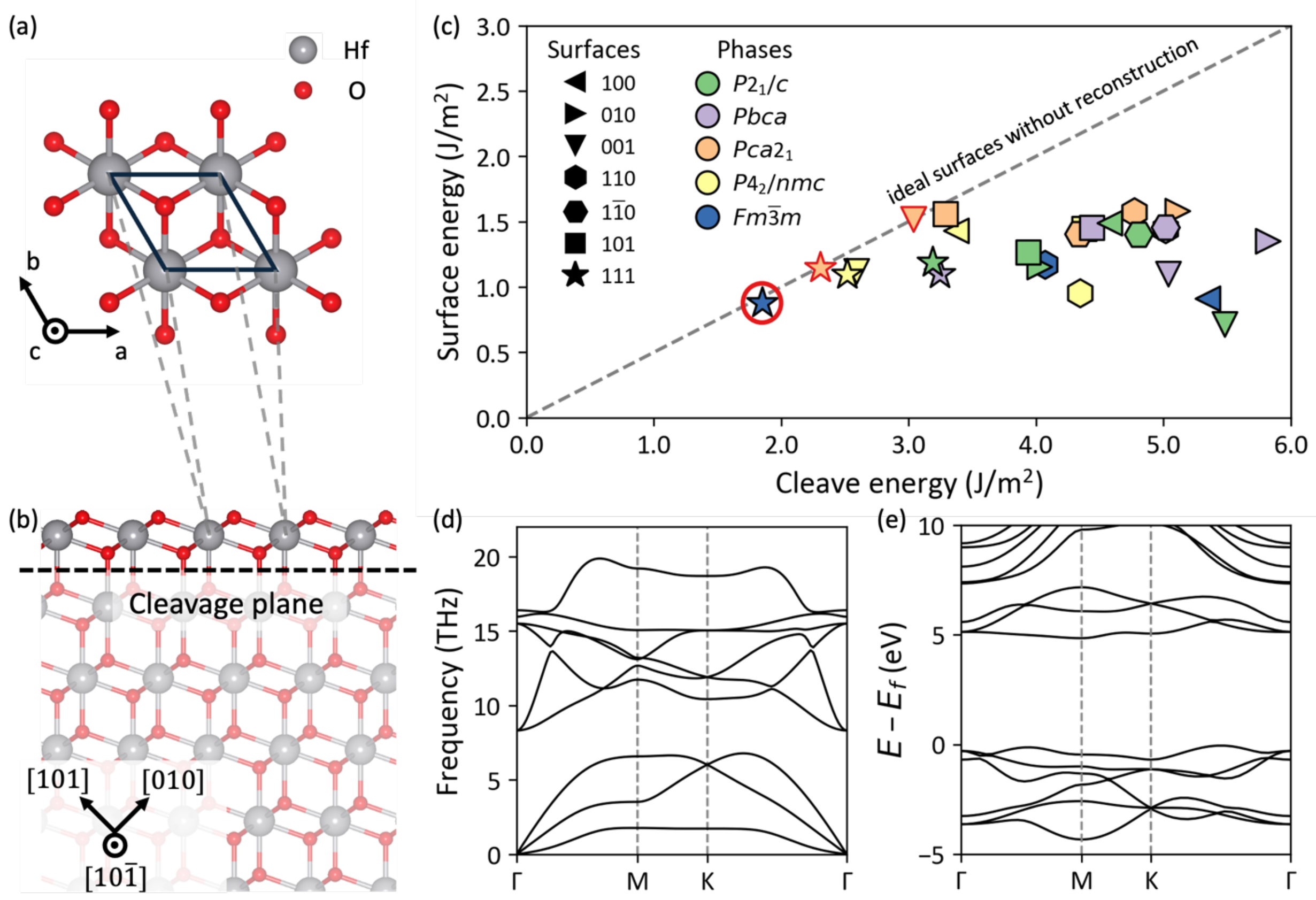

**Figure 1.** Crystal structure and stability of monolayer 1T-$HfO_2$. a) Top-view schematic of the atomic structure of monolayer 1T-$HfO_2$. b) Perspective of bulk cubic $HfO_2$ along the $[10\bar{1}]$ direction. c) Variation of cleavage energy and surface energy across different phases and surfaces of bulk $HfO_2$. Surfaces are distinguished by marker shape, and phases are differentiated by colors. 1T-$HfO_2$ can be viewed from the cleavage of the (111) surface of the cubic phase ($Fm\bar{3}m$), highlighted with a red circle. d) Phonon dispersion of monolayer 1T-$HfO_2$. e) Electronic band structure of monolayer 1T-$HfO_2$ calculated at the PBE level.

Next, we move to manipulating the stacking order in bilayer 1T-$HfO_2$. When considering only interlayer sliding, the bilayer stacking configurations can be classified as aligned and antialigned (Figure S2), with bulk $HfO_2$ adopting the aligned arrangement. We performed Berry phase calculations for both aligned and antialigned stacking bilayers to investigate the OOP polarization resulting only from sliding, as shown in Figure S2. For aligned stacking, interlayer sliding does not break the inversion symmetry, and thus, the net OOP polarization remains zero. In the antialigned case, symmetry analysis reveals that the polarization can be described by a non-harmonic trigonometric periodic function of the relative sliding displacement along the diagonal direction:[23]

$$P_{\perp} = P_1^{\mathrm{odd}}[2\sin(2\pi x) - \sin(4\pi x)], \quad (3)$$

where $P_{\perp}$ is the OOP polarization, $P_1^{\mathrm{odd}}$ denotes the first-order Fourier component of the OOP polarization, and $x$ is the relative sliding along the diagonal. The maximum OOP polarization occurs at AB/BA stacking orders and reaches 2.1 pC/m (equivalent to 1.1 μC/cm$^2$ in 3D), which is comparable to reported values in other sliding ferroelectrics such as $h$-BN (2.08 pC/m in 2D and 0.68 μC/cm$^2$ in 3D) and $MoS_2$ (0.97 pC/m in 2D and 0.52 μC/cm$^2$ in 3D).[21] However, this value remains significantly lower than that predicted in the bulk ferroelectric phase of hafnia with $Pca2_1$ space group (~50 μC/cm$^2$),[42] indicating that interlayer sliding alone is insufficient to induce substantial polarization in bilayer 1T-$HfO_2$. Moreover, introducing twisting into the antialigned stacking bilayer 1T-$HfO_2$ does not produce a net OOP polarization because the inversion symmetry is preserved.[43, 44]

We then investigate the influence of twisting aligned bilayer 1T-$HfO_2$ on the resulting OOP polarization. The twisted structures were generated by stacking two aligned monolayers with a relative rotation about a selected axis to create a moiré superlattice (MSL). Prior to running first-principles calculations, symmetry analysis provides critical insights into whether specific stacking configurations are capable of supporting OOP polarization. When the twist axis passes through Hf atomic sites, an Hf-axis MSL is formed (space group: $P321$), comprising three distinct domains corresponding to high-symmetry stacking orders: AA, AB, and AC, depicted in Figure 2b-d. Alternatively, when the twist axis passes through O atomic sites, an O-axis MSL is generated (space group: $P3$), also consisting of three domains corresponding to the same high-symmetry stacking configurations, as shown in Figure 2e-h.

Now we examine the differences between each domain (AA, AB, and AC) within the MSL and their corresponding high-symmetry stacking orders. Here, we use stacking orders to denote the ideal, high-symmetry bilayer structures, whereas stacking domains refer to the local structures within the MSL. Although all three high-symmetry stacking orders preserve the inversion symmetry and therefore prohibit the emergence of OOP polarization, the local domains within the MSL differ from these idealized stacking orders. Stacking domains can be regarded as slightly twisted variants derived from the high-symmetry stacking orders along the selected twist axis. For instance, in the Hf-axis MSL (Figure 2b-d), where the twist axis coincides with an inversion center, the twisting reduces the inversion symmetry to an in-plane twofold rotational symmetry, which forbids the appearance of OOP polarization. In contrast, in the O-axis MSL (Figure 2f-h), where the twist axis does not coincide with an inversion center, the inversion symmetry is lifted. The symmetry-breaking establishes the necessary conditions for the emergence of a net OOP polarization. Further discussion of the symmetry-driven mechanism is provided in Section S1 of the Supporting Information. Moreover, we find that interlayer sliding along the diagonal direction within the MSL is effectively equivalent to switching the twist axis in the aligned bilayer configuration, as illustrated in Figure S3 and discussed in Section S2 of the Supporting Information. This equivalence suggests that sliding of twisted bilayer 1T-$HfO_2$ can enable transitions between states that either forbid or permit OOP polarization. Notably, within this sliding-mediated switching mechanism, the sliding displacement direction is perpendicular to the polarization direction. Similar polarization switching driven by interlayer sliding has been reported in sliding ferroelectrics, such as *h*-BN, where relative sliding alters the point group from $D_{3h}$ (forbids OOP polarization) to $C_{3v}$ (permits OOP polarization), thereby enabling the emergence of a finite polarization.[21] We note that the above analysis is purely based on symmetry considerations and, therefore, is broadly applicable to all twisted bilayers adopting the 1T phase with aligned stacking configurations. Symmetry analyses by Ji *et al.*[45] and Xin *et al.*[46] have focused primarily on the role of interlayer sliding symmetry operations. Extending such symmetry frameworks to include both sliding and twist operations could provide a robust theoretical foundation for understanding the emergence of ferroelectricity in both sliding and moiré ferroelectric systems.

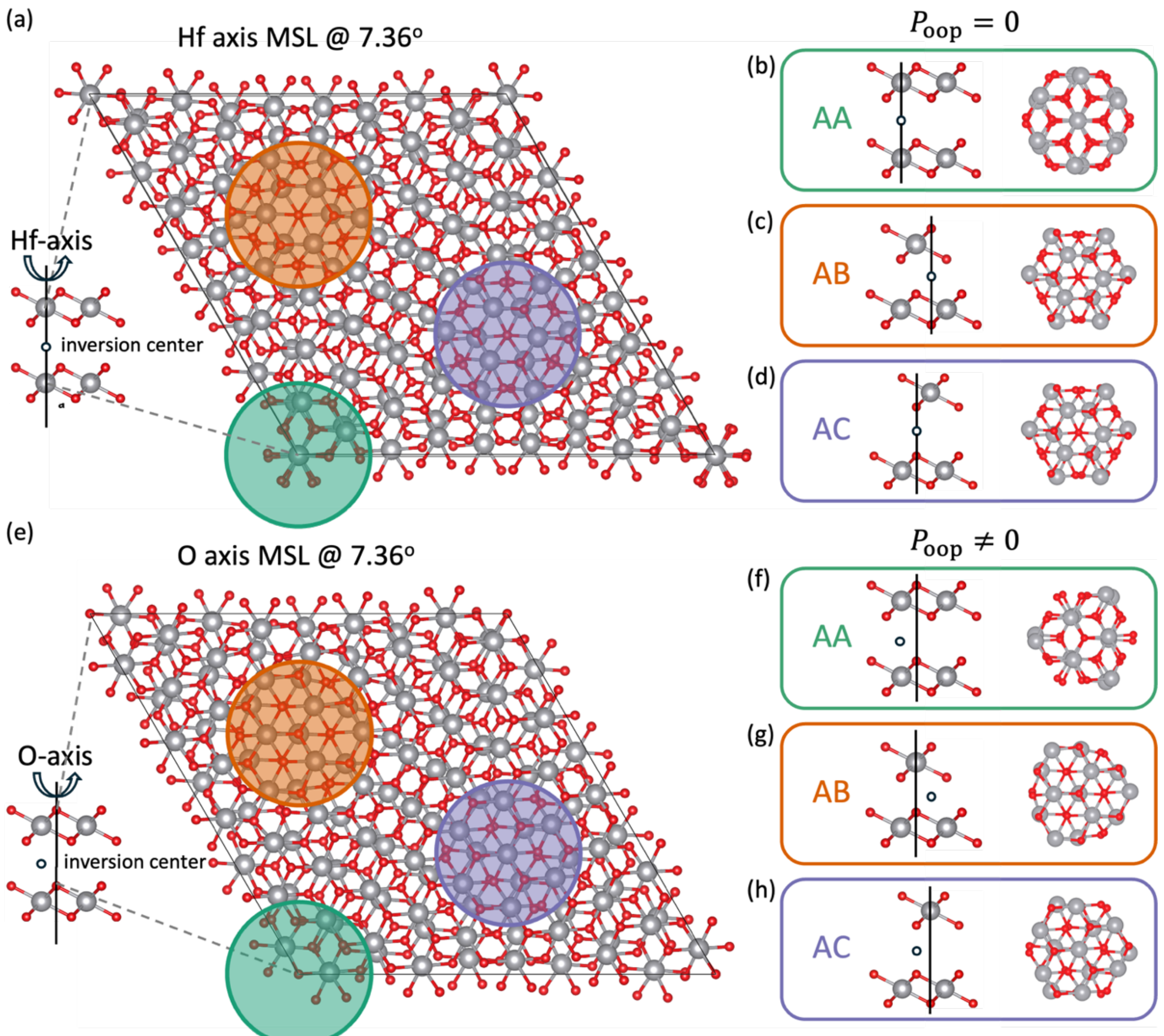

**Figure 2.** Crystal structure of MSLs. Top view of (a) Hf-axis MSL and (e) O-axis MSL twisted at 7.36°. b-d) and f-h) show the respective local stacking domains, which are distorted along different twisted axes and categorized by circles of different colors for Hf-axis MSL and O-axis MSL, respectively. The inversion center and twisted axis are shown as a black circle and line.

To quantify the OOP polarization in the MSLs, we use the OOP displacement of Hf atoms, defined as the displacement along the *z*-axis from the geometric center of the surrounding six oxygen atoms following structural relaxation. The number of atoms in the MSL increases super-exponentially as the twist-angle decreases, as shown in Figure S5. At a twist angle of 7.34°, the supercell contains 366 atoms, which is still tractable for DFT calculations, and the corresponding spatial distribution of polar displacements is presented in Figure 3a. The results reveal that polar displacements are predominantly localized within the AB domains for both the Hf-axis MSL and the O-axis MSL. We now turn to the question of what drives the concentration of polar displacements in the AB domains. To elucidate the energy landscape, the generalized stacking fault energy (GSFE) as a function of in-plane sliding was calculated,[47] as illustrated in Figure S6. The AB stacking order exhibits the strongest interlayer interactions among the three high-symmetry stacking orders, as evidenced by the GSFE profile. As a result, while the AA and AC stacking orders display electronic structures similar to that of monolayer 1T-$HfO_2$, indicating weak interlayer coupling, the AB stacking order exhibits noticeable band splitting, especially at the *K* point (Figure S7), indicating that the interlayer bonding characteristic deviates from vdW-like interactions.[30]

The enhanced interlayer interaction in the AB order is further reflected in the energy landscape and interlayer distances within the MSL. For AA and AC stacking orders, the energy landscape exhibits a single minimum, whereas AB stacking orders display two distinct local minima at different interlayer distances, namely bulk-like and vdW-like, as shown in Figure S8. The AB stacking order achieves the smallest equilibrium interlayer distance (vdW-like: 4.05 Å, bulk-like: 3.06 Å) compared to AA (4.73 Å) and AC (5.15 Å) stacking orders. Concurrently, the AB stacking order exhibits a slightly larger in-plane lattice constant (vdW-like: 3.25 Å, bulk-like: 3.38 Å) relative to AA and AC (both 3.23 Å). This behavior arises from the interplay between vdW interactions and the bulk-like structural characteristics inherent to the AB stacking order.[37] In particular, the cubic phase of $HfO_2$ adopts an ABCABC... stacking sequence, and the AB bulk-like stacking order can be interpreted as a bilayer cleaved from this bulk cubic structure.

We also computed the phonon band structures for all the three high-symmetry stacking configurations, which are shown in Figure S10. The AA, AC, and AB bulk-like stacking orders exhibit no imaginary frequencies throughout the Brillouin zone, indicating dynamical stability. In contrast, the AB vdW-like stacking order displays a softened acoustic out-of-plane flexural phonon (ZA) branch near the zone boundary, suggesting a potential lattice instability. To further investigate this instability, we visualized the phonon eigenvector corresponding to the soft mode at the M point—referred to as the $M_2^-$ mode in Figure S11. This mode is characterized by an antipolar displacement pattern of Hf atoms, which resembles the OOP polarization distribution observed in the AB domain in Figure 3a. Specifically, centered away from the bottom layer AB domain in the O-axis MSL, the first-nearest-neighbor Hf atoms are displaced downward, while the second-nearest neighbors are displaced upward, forming an alternating antipolar motif corresponding to the $M_2^-$ mode. Together with the stacking energetics, these observations indicate that the AB domain is particularly susceptible to soft-mode driven distortions and that this lattice relaxation locally condenses within the AB domain. As a result, the dominant lattice reconstruction in this moiré system is not merely in-plane strain accommodation or domain-size adjustment, but rather a soft-phonon driven enhancement of vertical ionic displacements that becomes strongly localized in specific stacking domains. Importantly, although the soft-mode eigenvector is antipolar, the moiré superlattice breaks the equivalence of local environments and weights them unevenly with respect to their distance from the twist axis, yielding a nonzero spatial average of the vertical displacements and thus a finite macroscopic OOP polarization. The OOP polarization is primarily ionic in origin, which is dominated by atomic displacements rather than interlayer charge transfer, so that the large displacement stabilized within AB domains naturally yields a sizable macroscopic polarization.

We also project the OOP polar displacements to the diagonal for the Hf-axis MSL and O-axis MSL twisted at 7.34° in Figure S12. Domains in the Hf-axis MSL, as predicted by symmetry analysis, preserve in-plane twofold rotational symmetry. Thus, even if polar displacements are concentrated at the AB domains, the dipoles cancel each other out for the bottom layer and the top layer, yielding zero net OOP polarization. In contrast, the O-axis MSL, lacking symmetry elements that forbid OOP polarization, exhibits a net polarization across the twisted bilayer, as schematically illustrated in Figure S13. We also investigate the twist-angle dependence of the net OOP polarization, as shown in Figure S14. Calculations up to a twist angle of 6°, corresponding to an MSL containing 546 atoms, reveal that the net OOP polarization increases monotonically as the twist angle decreases up to 6°. The enhancement is attributed to increased lattice reconstruction at lower twist angles. We anticipate that at even smaller twist angles, the net polarization may eventually decrease due to competition between interlayer interaction and lattice mismatch, a phenomenon previously discussed in the context of twisted bilayer $MoS_2$.[48] Also, owing to the 60° rotational symmetry of the bilayer lattice, the structural distortions and the associated polarization are periodic functions of the twist angle. Specifically, twist angles of $\theta$, $n\times60^\circ - \theta$, and $n\times60^\circ + \theta$ (where n is an integer) yield symmetry-equivalent configurations and therefore identical magnitudes of polarization. The present calculations up to 6° thus represent the small-angle branch within one such symmetry period.

Furthermore, as discussed in Figure S3 and Section S2 of the Supporting Information, relative interlayer sliding effectively modifies the twist axis of the bilayers, providing a mechanism for polarization switching. To validate this, we computed the total energy and net OOP polarization for sliding the MSL twisted at 7.34° along the diagonal direction of the MSL. The local dipole moments were estimated using the Born effective charges and the off-center displacements.[23, 30] The calculated double-well energy landscape is shown in Figure 3b, where the Hf-axis MSL corresponds to the nonpolar state, while the O-axis MSL represents the polar state. The net OOP polarization associated with the O-axis MSL can be understood as resulting from a relative interlayer sliding of one-third along the diagonal direction with respect to the Hf-axis MSL. Notably, the O-axis MSL also corresponds to the lower-energy state.

The polarized state is expected to be robust against depolarization fields. As the amplitude of the polar displacements is increased, the polarized bilayer structure consistently has a lower energy than its non-polar counterpart, as shown in Figure S15. Previous experimental studies have demonstrated that stacking orders in layered materials can be interconverted by applying a vertical electric field, which drives relative in-plane motion between adjacent layers.[13, 14, 16, 17] The same mechanism is expected to operate here: to align the external OOP electric field, the lateral alignment of the twisted bilayer will switch by a relative shift along the diagonal of the MSL. The interlayer interaction between twisted bilayers is primarily the vdW force, lowering the energetic cost for sliding. This weak coupling therefore allows reversible transitions between the two polarized O-axis MSLs. We attribute the switching mechanism to bilayer sliding rather than domain dynamics.[32] Here, we clearly show a bistable polar state, suggesting that bilayer sliding is a feasible switching mechanism. The associated energy barrier for this transition is ~8 meV/f.u., and the expected coercive field for coherent switching is ~0.2 V/nm (see Section S3 of the Supporting Information).[16] This value is comparable to those reported in other sliding ferroelectrics, such as *h*-BN (~0.1 V/nm) and $WS_2$ (~0.3 V/nm).[13, 14, 16, 49] This highlights its potential as a scalable ferroelectric platform capable of retaining low coercive fields even at the nanoscale, which suggests that twisted bilayer 1T-$HfO_2$ could be used in ultrathin devices.

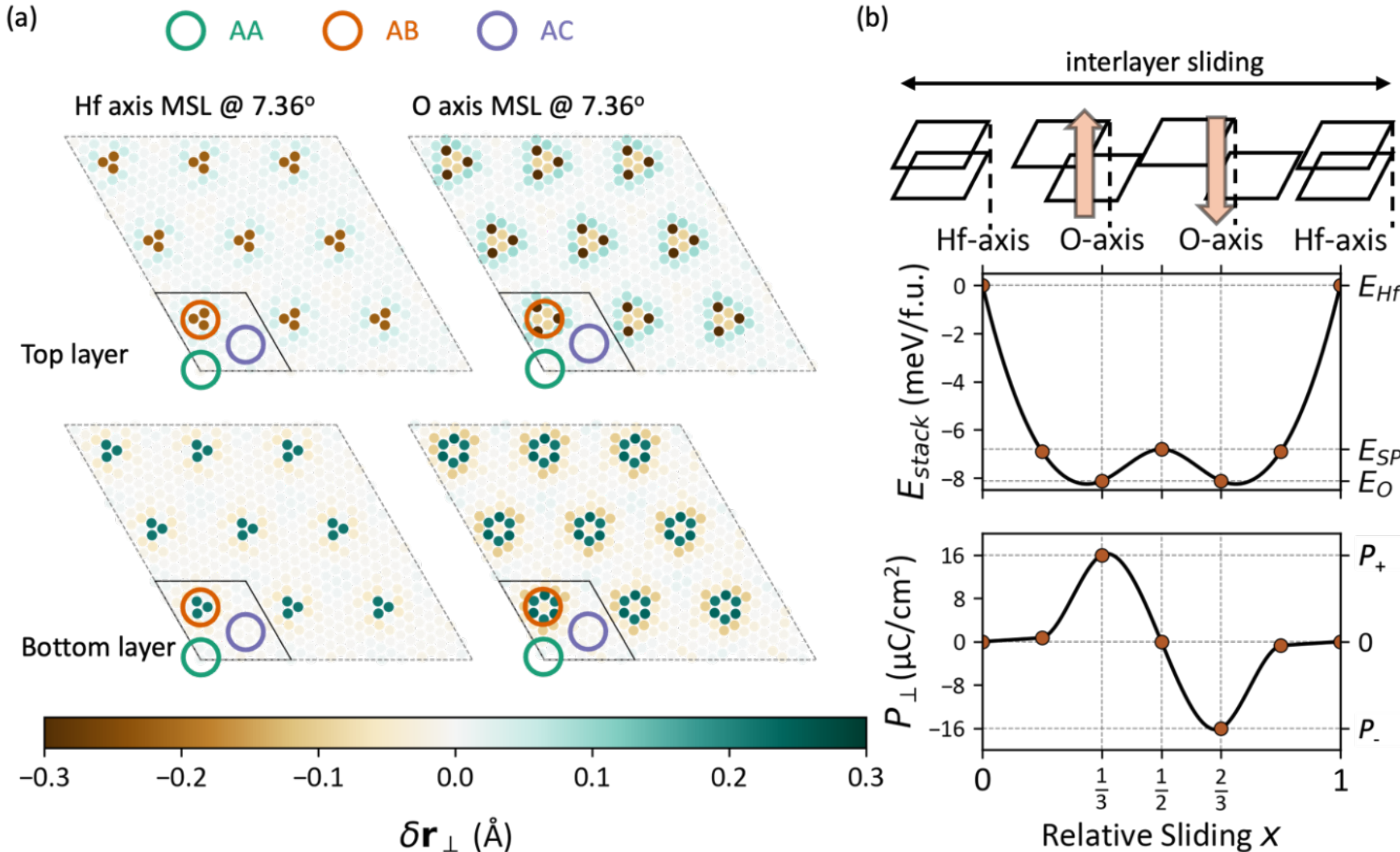

**Figure 3.** Ferroelectricity in twisted bilayers 1T-$HfO_2$ with the twisted angle of 7.36°. a) OOP polar displacements mapping. Domains are categorized by differently colored circles. b) Stacking energy and OOP polarization in twisted bilayer 1T-$HfO_2$ as a function of relative sliding along the MSL diagonal.

In summary, we have systematically investigated the structural and ferroelectric properties of twisted bilayer 1T-$HfO_2$. We predict that it should be feasible to obtain monolayer 1T-$HfO_2$ by cleaving it from the bulk cubic phase. Calculations of surface and cleavage energies, energy above the convex hull, and phonon dispersion collectively indicate the thermodynamic and dynamic stability of monolayer 1T-$HfO_2$. Through symmetry analysis and DFT calculations, we demonstrated that the relative twisting of bilayers breaks inversion symmetry, enabling the emergence of a net OOP polarization. Furthermore, we identified that the enhanced interlayer interaction at the AB stacking domains leads to a significant concentration of polar displacements localized at these sites. For MSLs that allow net polarization (i.e., O-axis MSLs), the net OOP polarization can reach ~16 μC/cm² at a twist angle of 7.34°, and the polarization is switchable through interlayer sliding. The predicted polarization in bilayer MSLs of 1T-$HfO_2$ is approaching that of the bulk polar phase of $HfO_2$, and has a significantly lower coercive field of 0.2 V/nm, compared to 0.4 V/nm in bulk $HfO_2$.[50] These observations highlight the potential of twisted bilayer 1T-$HfO_2$ as a promising platform for achieving robust, scalable OOP polarization in the low dimension, offering significant opportunities for integration into next-generation nanoscale electronic devices.

METHODS:

*Density-functional theory (DFT) calculations*: DFT calculations were performed using projector-augmented wave (PAW) potentials, as implemented in the Vienna ab initio simulation package (VASP).[51] The Perdew–Burke–Ernzerhof (PBE) functional within the generalized gradient approximation (GGA) was used to describe exchange-correlation interactions. A *k*-point spacing of 0.03 $Å^{-1}$ and an energy cutoff of 500 eV were selected for both structure optimization and static calculations. The convergence criteria for structure relaxations were set to $10^{-6}$ eV and $10^{-4}$ eV/Å for energy and forces, respectively. A slab with a vacuum of 15 Å was used for the bilayers. A dipole correction term was used to remove the electric field in the vacuum region.[52, 53] To account for the vdW interlayer interactions, the DFT-D3 method with Becke-Johnson damping function was employed.[54] Structural stability was evaluated through phonon calculations using the finite displacement method, which is implemented in phonopy.[55] A 6 × 6 × 1 supercell (108 atoms) was constructed for force-constant calculations, with atomic displacements of 0.01 Å, and the convergence criterion was set to $10^{-8}$ eV. For MSL calculations, the convergence criteria were reduced to $10^{-5}$ eV and $10^{-2}$ eV/Å for energy and forces, respectively, following Lee *et al.*[30] The local dipole moment, $P^i$, of each unit cell was defined as:[56]

$$P^i = \frac{\mathrm{e}}{\Omega_\mathrm{c}} \sum_\alpha w_\alpha \mathbf{Z}^*_\alpha \mathbf{u}^i_\alpha \tag{4}$$

Here, e is the electron charge, $\mathbf{Z}^*_\alpha$ is the Born effective charge tensor, $\mathbf{u}^i_\alpha$ are displacements of the atom α away from the centroid of the octahedron, $w_\alpha$ is the weight factor of the atom α, $i$ is the index of the unit cell, and $\Omega_\mathrm{c}$ is the volume of the unit cell to evaluate the local dipole moment. The net polarization is calculated by summing up all the local dipole moments.

ACKNOWLEDGEMENTS: This work was primarily supported by the National Science Foundation (NSF) through awards DMR-2145797 (J.H., P.O., G.Y.J., R.M.), DMR-2122070 (J.H., G.R., R.M.), and DMR-2419172 (X.L., X.X.). Computational resources were provided through allocation DMR160007 from the Advanced Cyberinfrastructure Coordination Ecosystem: Services & Support (ACCESS) program, supported by NSF grants #2138259, #2138286, #2138307, #2137603, and #2138296. L.Y. is supported by the Office of Science of the U.S. Department of Energy under Contract No. DE-SC0026312.

ASSOCIATED CONTENT: The data that support the findings of this study are available in Twist-induced Out-of-plane Ferroelectricity in Bilayer Hafnia at https://zenodo.org/records/18662851.[57]